\documentclass[aps,prb,twocolumn,amsmath,amssymb,eqsecnum,superscriptaddress,floatfix]{revtex4}
\usepackage{graphicx}

\begin{document}
\title{Fermion Chern Simons Theory of Hierarchical Fractional Quantum Hall States}

\author{Ana L\'opez}
\altaffiliation{Permanent address: Centro At{\'o}mico Bariloche,
(8400) S.\ C.\ de Bariloche,
R{\'\i}o Negro, Argentina}
\affiliation{Department of Physics, Theoretical Physics, Oxford
University, 1 Kebble Road, Oxford OX2 3NP, UK}
\author{Eduardo Fradkin}
\affiliation{Department of Physics, University of Illinois at Urbana-Champaign, 1110 W. Green St., Urbana IL 61801-3080, USA}

\date{\today}
\begin{abstract}
We present an  effective Chern-Simons theory for the bulk fully 
polarized fractional
quantum Hall (FQH) hierarchical states constructed as  daughters 
of general states of
the Jain series, {\it i. e.\/} as FQH states of the quasi-particles or
quasi-holes of Jain  states. We discuss the stability of these
new states and present two reasonable stability criteria. We discuss
the theory of their edge 
states which follows naturally from
this bulk theory. We construct the operators that create elementary
excitations, and discuss the scaling behavior of the tunneling
conductance in different situations. Under the assumption that the
edge states of these fully  polarized hierarchical states are
unreconstructed and unresolved, 
we find that the differential conductance $G$ for tunneling of
electrons from a Fermi liquid into {\em any} hierarchical Jain FQH
states  has the scaling behavior $G\sim V^\alpha$ with the universal exponent
$\alpha=1/\nu$, where $\nu$ is the filling fraction of the
hierarchical state. Finally, we  explore
alternative ways of constructing FQH states with the same filling 
fractions as partially polarized states, and conclude that this is not
possible within our approach.
\end{abstract}


\maketitle

 In 1983 Laughlin [\onlinecite{laughlin}] proposed his celebrated wave functions as
an explanation of the fractional quantum Hall effect (FQHE) for a
two-dimensional electron gas (2DEG) with filling factors $\nu=1/m$,
with $m$ an odd integer. Shortly after that, hierarchical
generalizations of these fully polarized states were also proposed for
arbitrary filling factors with odd-denominator
fractions~[\onlinecite{halperin,haldane}]. The basic idea behind these
Halperin-Haldane hierarchical states is that, at filling factors away from $\nu=1/m$,
there is a finite density of excitations of a primary (Laughlin)
fractional quantum Hall (FQH) state, that are quasi-holes with
fractional charge and fractional statistics, which themselves may also
condense into a new incompressible FQH state. Thus, for instance the
FQH state at $\nu=2/5$ is viewed as a FQH state of quasi-holes of the
$\nu=1/3$ state, a ``condensate".  Away from these filling factors,
there is a finite density of these new excitations which in turn may
also condense into yet another incompressible FQH state, and so on. 
The bosonic Chern-Simons field theory of the Laughlin states~[\onlinecite{zhk}] describes the 
FQH states as the Bose condensation of a (bosonic) field. In this picture
 the Haldane-Halperin hierarchical states are described by the
condensation of vortex excitations on the primary condensate.~[\onlinecite{wen}]
 
Thus, Halperin-Haldane hierarchical FQH states have a built-in nested structure, much
reminiscent to that of a Russian doll, in which states higher in the
hierarchy are supported by lower states and so forth. Implicit in
this construction is the assumption that, as the magnetic field or
the density are varied, their excitations remain well defined even
when their densities are relatively high (as measured by their
effective filling factors). Besides, underlying lower states remain
stable and inert as their excitations condense into yet new higher
states. Thus, although the Halperin-Haldane hierarchy offers a simple way to construct
states for arbitrary filling, it is a crude model whose assumptions
may not obviously hold.  Furthermore, it is a necessary consequence
of the assumptions of this hierarchical construction that the Hilbert
spaces supported by these states have many conserved currents, one
for each level of the hierarchy, even though the underlying 2DEG {\it
a priori} has only one conserved current, the charge current, as
required by charge conservation. Thus even if the Halperin-Haldane hierarchical states
may be sound in the low energy limit, at the microscopic level most of
these additional conservation laws must be violated. Hence, for this
hierarchy to work it is necessary to suppress the processes which
violate the extra conservation laws. It also follows from
these constructions that, as the physical edges of the 2DEG are
approached, the intricate structures of these states is progressively
revealed as one layer after another of the state structure gets
peeled-off~[\onlinecite{wz,wen}]. This nested structure also determines many
qualitative features of the phase diagram of the 2DEG in the regime in
which only incompressible fluids and insulating states are present~[\onlinecite{kivelson92}].

An alternative construction, based on the notion of composite
fermions, was proposed by Jain [\onlinecite{jain89a}]. This approach yields
stable FQH states for filling factors on the sequences $1/\nu=2n+1/p$,
where $n$ and $p$ are integers ($n\geq 0$). The FQH states on the Jain sequences are the most prominent, and hence
the most stable, of the experimentally observed FQH states. In these families of
states, within a mean-field picture, the composite fermions fill up a
finite number $|p|$ of Landau levels of a partially screened magnetic
field. Beyond mean field theory fluctuation effects, embodied by an
effective Chern-Simons gauge field~[\onlinecite{lopez91}], turn the mean
field composite fermions into excitations with fractional charge and
fractional statistics~[\onlinecite{lopez91}]. It is also possible to construct a 
hierarchy of FQH states using composite fermions~[\onlinecite{quinn97,mani96}]. 

Although the universal physical properties of the FQH states, {\it i.e.\/} the charge and statistics 
of the excitations and the ground state degeneracy,  derived from both constructions are in fact
 identical~[\onlinecite{zhk,wz,lopez91}],  in practice the nature of the 
approximations that are made are different in each approach, {\it i. e.\/} 
condensation of a composite boson or filling up Landau levels of composite fermions. 
As a result, there are significant quantitative
differences in their predictions of energy gaps and other
non-universal but physically important properties.  
Besides, as we already discussed, it is an experimental fact
that the states in the main Jain sequence are  the most stable
fully polarized states, while within the Haldane-Halperin approach,
only the Laughlin states would be the naturally more stable. Why this is true is not entirely clear
theoretically since these theories (bosonic or fermionic) do not have a small expansion
parameter and the corrections to the mean field results may be large
(and in practice are large). Typically mean field theory overestimates the energy
gaps of the Jain states by about an order of magnitude (compared with
numerical results for small systems). Also, the flux attachment transformation, 
central to both constructions, is a local operation in space and as such it involves a 
large amount of Landau level mixing. 
While this is not a problem for the determination of the universal data of FQH states, 
it does have a large effect on energy gaps and similar quantities. In particular the mean field 
theory yields gaps whose magnitude is controlled by the effective magnetic field acting on 
the composite fermions and not by the physical energy scale, the Coulomb energy at the 
magnetic length. Calculations beyond mean field theory may solve in the end some of these problems
~[\onlinecite{shankar97,murthy02,shankar02}]. Numerical calculations with a relatively small number 
of electrons, presumably with significant and yet poorly understood finite size effects, 
support the conclusion that the Jain states are indeed more stable and that their 
stability diminishes with the order of the Jain sequence~[\onlinecite{quinn00}] . Thus, at least 
qualitatively, the stability of the
experimentally observed states, as measured by the width of their respective Hall
conductivity plateaus, follows the progression of the Jain sequences, 
rather than the Haldane-Halperin hierarchy~[\onlinecite{halperin,haldane}]. 

As we noted above, the Halperin-Haldane hierarchy leads to an effective low-energy theory which involves a 
number of conserved currents, one per level of the hierarchy, which thus grows with the level. 
This is actually a feature inherent to all hierarchical constructions, either fermionic or bosonic. 
In the $K$-matrix form of the effective theory, this feature is encoded in the rank of the $K$-matrix 
{\em and} on the number of fundamental quasiparticles, both of which grow with the level of the 
hierarchy[\onlinecite{wz,wen}]. It was noted by Haldane[\onlinecite{haldane-t}] that in most cases 
the resulting $K$-matrix has ``null vectors", representing neutral particles with ``zero statistics" ({\it i.e.\/} without chirality), 
and proposed that such particles need not be conserved thus leading to an instability of these states. 
He then went on to propose that the states whose $K$-matrices do not have null vectors are the only stable states and 
coined the term $T$-stability to describe this criterion. 

Motivated by the question of whether these additional conservation laws are actually needed to understand the FQH 
states we have recently constructed a theory for all FQH states in the Jain sequences which requires the existence 
of only one conserved current, the charge current, and only one fundamental quasiparticle[\onlinecite{lopez99,lopez01}]. 
A key ingredient of this theory is a generalization of the flux attachment transformation 
which is consistent with the requirements resulting from topological and gauge invariance applicable for a 2DEG on 
a closed surface. Unlike the standard hierarchy, the resulting effective Chern-Simons theory has several components 
and it is characterized by a $K$-matrix whose rank is the same for all the sequences. Moreover the resulting theory 
is automatically $T$-stable in the sense of Haldane.
  
Recently,  Pan {\it et al.\/} [\onlinecite{pan03}]  have carried out a set of experiments on high
mobility samples at very low temperatures in which they observed the
FQHE in the lowest Landau level at filling fractions not included in the
quantum Hall (Jain) series.
A deep minimum in $\rho_{xx}$, as well as a respectable plateau in $\rho_{xy}$, was
observed for  $\nu=4/11$ and $5/13$, whereas for 
$\nu=6/17,4/13,5/17$, and $7/11$, the minimum was not as pronounced,
and in fact no plateau in $\rho_{xy}$ was actually observed for these filling fractions. 
The state at $7/11$ was observed earlier by Goldman and Shayegan [\onlinecite{goldman90}].
A respectable minimum in $\rho_{xx}$ was also seen for $\nu=3/8$, but without any evidence 
for a plateau in $\rho_{xy}$.
In the case of  the best defined of these new states, with $\nu=4/11$,
Pan {\it et al.\/} [\onlinecite{pan03}] suggested that it is a fully polarized  FQH state.
They further proposed that the states they observed are
evidence of a FQHE of composite fermions. 

If this interpretation is correct, these are the first truly hierarchical FQH states observed 
to date, {\it i. e.\/} FQH states of the physical excitations of a primary Jain state. However, 
we should note that there are other possible 
mechanisms to get an incompressible FQH fluid, at least for some of the filling fractions they observed. 
Finite size diagonalization of small clusters of electrons 
[\onlinecite{jain02a,jain03b,note}] suggest that some of the observed states, such as 
the one at $\nu=4/11$, should be at most partially polarized. This result is, however, at variance 
with the experiments of Ref. [\onlinecite{pan03}] that are consistent with a fully polarized state.  
On the other hand, 
a number of authors [\onlinecite{scajain,quinn03a,quinn03b}] have also proposed that at least for some 
of the observed fractions, for example $\nu=3/8$, the ground state may be a paired quantum Hall 
state, with pairing of the excitations of the Laughlin $\nu=1/3$ FQH state~[\onlinecite{foot}]. It is 
clear that in principle there may be two or more competing phases and that whichever state is 
observed may depend on subtle microscopic details.

In this paper we will take the point of view that the observed states
are indeed hierarchical Jain states. In other words, these new states
are the result of the condensation of the {\em physical
quasi-particles and quasi-holes} of the primary Jain states
into  new Jain-like FQH states. We  will use here the approach we introduced in Refs. [\onlinecite{lopez99,lopez01}] 
to construct these new states as hierarchical Jain states, and to derive an effective Chern-Simons field theory for these states, 
demanding the existence of only the minimum number of necessary conservation 
laws and compatible with the consistency requirements when the 2DEG is placed on a closed surface. 
We will show that these states can be (locally) stable and that they satisfy the requirements for $T$-stability.  
We will also show that these new states are organized more naturally as Jain hierarchical states rather than along 
the lines suggested by the standard Haldane-Halperin hierarchy. We will use the resulting bulk theory to find 
an effective theory of an 
unreconstructed unresolved edge and to compute the tunneling exponent for electrons into these states. We find that 
it is  equal to $1/\nu$ {\em for all} systems with unreconstructed, 
unresolved ({\it i. e.\/} ``sharp") edge states within the 
Jain series as well as for all of their hierarchical descendants. This result, which we found earlier to hold for the primary Jain states [\onlinecite{lopez99}], is consistent with all the presently 
available experimental data on the tunneling exponent [\onlinecite{chang96,grayson98,hilke01,hilke03}]. It 
suggests that this dependence of the exponent on the filling factor is a generic property of all clean unreconstructed unresolved edges [\onlinecite{foot-edge}].
In contrast, it is well known that if the edges are resolved, 
whether they are clean or not, the exponent has a more complex dependence on 
the filling fraction [\onlinecite{weniv,kaneiv,kane-fisher-polchinski,shytov98}]. 

The paper is organized as follows. In Section \ref{new} we introduce a
generalization of the fermionic Chern-Simons theory of the FQHE for
fully polarized systems along the lines of Refs. ~[\onlinecite{lopez91,lopez99}], and use it to construct these states, 
to compute their degeneracy on a torus, and to determine the quantum
numbers of their quasi-hole and quasi-particle excitations. We also show that, contrary to the 
general expectations derived from the standard hierarchy[\onlinecite{haldane-t}] these states are $T$-stable. 
In doing so we will assume that quasi-particles and quasi-holes of the 
primary Jain states have simple and short range interactions. Under these assumptions it 
is possible to give a simple (perhaps naive) criterion for the stability of these states, 
which is discussed in Section \ref{criterion}. 
In Section \ref{partial} we study the possibility of these states
being realized as partially polarized primary FQH state of electrons, 
and conclude that this option is not generally 
feasible within our approach,  for all the filling fractions that have been observed.
Finally in Section \ref{edges} we derive an effective theory for the
edge states for these 
new hierarchical states, and calculate the electron edge tunneling 
exponents. 

\section{Fully polarized hierarchical Jain states}
\label{new}

The elementary excitations of all FQH states, including those on the
principal Jain sequences with filling factor $1/\nu=2n+1/p$,  are
quasi-holes with fractional charge and fractional statistics
~[\onlinecite{laughlin,halperin,zhk,lopez91,wen}]. Here we will take as the
starting point,  the
effective theory for the states in the Jain sequences developed in our
earlier work [\onlinecite{lopez99}], which follows the framework and notation
of Wen [\onlinecite{wen}]. In this work the elementary excitations of these states
are described by a set of currents $j_{qp}^\mu$, the world-lines of a
set of (composite) fermions,  which are coupled to a statistical gauge
field $a_\mu$ and to a hydrodynamic gauge field $b_\mu$ through the
effective action
\begin{eqnarray}
{\cal L} &=& {\frac {p}{4\pi}} \epsilon_{\mu\nu\lambda} a_\mu \partial_\nu 
a_\lambda + 
{\frac {1}{2\pi}} \epsilon_{\mu\nu\lambda} a_\mu \partial_\nu
b_\lambda   \nonumber \\
&-&{\frac {2n}{4\pi}} \epsilon_{\mu\nu\lambda} b_\mu \partial_\nu b_\lambda + 
{\frac {1}{2\pi}} \epsilon_{\mu\nu\lambda} b_\mu \partial_\nu A_\lambda +
a_\mu j^\mu_{qp}
\label{L}
\end{eqnarray}
where $A_\mu$ is an external electromagnetic perturbation. This
effective action reproduces all the universal data of the Jain states:
the $2np+1$-fold ground state degeneracy on the torus, the
quasi-particle fractional charge $e/(2np+1)$ and fractional statistics
$2n\pi/(2np+1)$ (measured relative to fermions)~[\onlinecite{lopez99}].

Away from these precise filling factors, the system has a finite
density of quasi-particles (or quasi-holes) which, due to their residual
interactions, can condense in a new FQH state in the  field
$a_\mu$. Thus, we will assume that the underlying Jain state remains
stable even in the presence of a finite density of its elementary excitations.
We can then apply the same procedure for these quasi-particles as we did for the original electrons, and
attach an even number $2 n_1$  of flux quanta to each
quasi-particle which  ensures that  their
statistics remains  unchanged, resulting in the quasi-particle Lagrangian:
\begin{eqnarray}
{\cal L}_{qp}&=& a_\mu j^\mu_{qp} \rightarrow 
a_\mu j^\mu_{qp}+ c_\mu j^\mu_{qp}+
{\frac {1}{2\pi}} \epsilon_{\mu\nu\lambda} c_\mu \partial_\nu d_\lambda  \nonumber \\
&-&{\frac {2 n_1}{4\pi}} \epsilon_{\mu\nu\lambda} d_\mu \partial_\nu
d_\lambda 
\label{Lqp1}
\end{eqnarray}
where we have written the conserved quasi-particle current as 
$j^\mu_{qp}=
{\frac {1}{2\pi}} \epsilon_{\mu\nu\lambda} \partial_\nu d_\lambda$.
Upon integrating out the quasi-particles, within a mean field theory in which
they fill up $|p_1|$ Landau levels of 
the effective field $a_\mu+ <c_\mu>$, the effective action becomes
\begin{eqnarray}
{\cal L}_{qp}&=&  {\frac {p_1}{4\pi}} \epsilon_{\mu\nu\lambda} c_\mu \partial_\nu c_\lambda  +
{\frac {1}{2\pi}} \epsilon_{\mu\nu\lambda} (c_\mu +a_\mu) \partial_\nu
d_\lambda  \nonumber \\
&-&{\frac {2 n_1}{4\pi}} \epsilon_{\mu\nu\lambda} d_\mu \partial_\nu d_\lambda 
\label{Lqp2}
\end{eqnarray}
where $c_\mu$ are the fluctuations about the mean field $\langle c_\mu \rangle$. Integrating out
$c_\mu$ and $d_\mu$  we find that the contribution of the quasi-particles to the total action is 
${\frac {\nu_1}{4\pi}} \epsilon_{\mu\nu\lambda} a_\mu \partial_\nu a_\lambda $, with  
\begin{equation}
\frac{1}{\nu_1}= 
2 n_1 + \frac{1}{p_1}
\label{nu1}
\end{equation}
 {\it i.e.\/} the quasi-particles condense in a Jain-like FQH state with
filling fraction $\nu_1$. So far, we have used the term quasi-particle
to name the elementary excitations of the Jain states, independently
of the sign of their charge. Notice, however, that if we take into
account this sign, inter-changing quasi-particles (negative charge) by
quasi-holes (positive charge)  will only change the sign of $\nu_1$. Thus, $\nu_1<0$
corresponds to a FQH state of quasi-holes and $\nu_1>0$ is a FQH state
of quasi-particles (or quasi-electrons). Hence there is no restriction on the sign of either $n_1$ or $p_1$. In contrast, we still have the constraint $n>0$ and that the {\em total filling factor} is positive. 

We can now collect the results of Eqs.(\ref{L})-(\ref{Lqp2}) in the  more 
compact $K$-matrix form~[\onlinecite{wen}].  In this representation, the effective Lagrangian 
 for a sequence of states whose ``primary" states are the Jain ones at
$1/\nu=2n+1/p$ is :
\begin{eqnarray}
{\cal L} &= &
{\frac {1}{4\pi}} K_{IJ} \epsilon_{\mu\nu\lambda} a_\mu^I \partial_\nu a_\lambda^J  +
{\frac {1}{2\pi}} t_I \epsilon_{\mu\nu\lambda} a_\mu^I \partial_\nu A_\lambda
 + \ell_I a_\mu^I j^\mu_{qp,1}\nonumber \\
 &&
 \label{eq:Ljain}
\end{eqnarray}
where the coupling constant matrix is
\begin{eqnarray}
K =  
\begin{pmatrix}
-2n  & 1  & 0  &    0 \\
  1 & p  & 1   &   0 \\
  0 & 1 &  -2 n_1  & 1 \\
  0 & 0 &  1  &    p_1
\end{pmatrix}
\label{K}
\end{eqnarray}
and the indices $I,J= 1,..,4$. 
We have defined the gauge fields   $a_\mu^1= b_\mu$,  
$a_\mu^2= a_\mu, a_\mu^3= d_\mu, a_\mu^4= c_\mu$, the charge vector
$t=(1,0,0,0)$, and  the flux vector $\ell=(0,0,0,1)$. $j^\mu_{qp,1}$ is
the quasi-particle current corresponding to the excitations above the
hierarchical state. Therefore, the  general set of allowed
quasiparticle excitations on top of this hierarchical state 
can be represented by a vector $m=k \ell=(0,0,0,k)$ where $k \in {\mathbb Z}$.

It follows that the filling fraction of these states is~[\onlinecite{wen}]
\begin{eqnarray}
\nu = |t^T K^{-1} t | = 
\frac {p (2 n_1 p_1 +1) + p_1}{\det K}
 \label{newnu}
\end{eqnarray}
or, equivalently
\begin{equation}
\frac{1}{\nu}=2n+\frac{1}{p+\nu_1}
\label{newnu2}
\end{equation}
where $\nu_1$ is given in Eq.(\ref{nu1}). It is straightforward to
check that these sequences include the series of FQH hierarchical
states  proposed earlier on in Ref. [\onlinecite{quinn97}].

In Eq. (\ref{K}), the quantity
\begin{equation}
|\det K| = |(2n_1 p_1 +1) (2np+1) +2n p_1|
\label{degeneracy}
\end{equation}
is the ground state degeneracy of these states on a torus. This result agrees with the standard hierarchy even though, 
as we shall see, the spectrum is not quite the same. It is also easy to check that, provided $|\det K|$ remains finite, 
the excitation spectrum has
a finite energy gap $E_G$ which in mean field theory is 
\begin{equation}
E_G=\displaystyle{\frac{\hbar \omega_c}{|\det K|}}
\label{gap}
\end{equation}
We expect that this mean field result is an over-estimate of the size of the real gap. 
As it is usual in these type of approximations~[\onlinecite{jain89a,lopez91,hlr}] the mean field gap is just the 
cyclotron gap of the composite fermions and it depends on the bare mass (in this case of the quasi-particles 
or quasi-holes which are condensing in the FQH state). Here too, on  physical grounds  one expects that the 
cyclotron scale should be replaced by the appropriate Coulomb interaction energy of two quasi-particles 
(or quasi-holes) at a distance of the order of the magnetic length. Thus, due to the short-distance 
structure of these excitations~[\onlinecite{gap-estimate}],
we expect that quasi-particle and quasi-hole gaps will 
be appreciably reduced from the mean-field theory prediction.

The quantum numbers of the quasi-holes of the new states are  
\begin{eqnarray}
Q_{qh} =e t^T K^{-1} \ell={\frac  {e}{\det K}}
\label{eq:newQ}
\end{eqnarray}
which is their charge, and 
\begin{eqnarray}
\frac {\theta_{qh}}{\pi} = \ell^T K^{-1} \ell =\frac {2n +2 n_1 (2np+1)}{\det K}
\label{eq:newtheta}
\end{eqnarray}
for their fractional statistics (measured from fermions). Note that there is only one 
quasi-particle (or quasi-hole) in our construction.
In contrast, the Haldane-Halperin hierarchy predicts the existence of a number
 of well defined 
distinct stable quasi-particles whose number given by the rank of the
matrix $K$,  depends on the  level of the hierarchy~[\onlinecite{wen}]. 
In particular, and in contrast to the requirements of the
Haldane-Halperin hierarchical construction, 
in our  theory there is one and only one conserved current with
respect to the electromagnetic 
field.

An interesting consequence of this fact is that, in contrast with what
happens in the case of the Haldane-Halperin hierarchical states, all the states described
by our approach are $T$-stable.  According to Haldane~[\onlinecite{haldane-t}], 
an Abelian Quantum Hall theory is $T$-unstable if the  effective $K$-matrix theory has
quasiparticles labeled by  vectors $m$ that satisfy $m K^{-1} m=0$ and
$t K^{-1} m=0$. An important caveat is that the null vectors must belong to the subspace spanned by the $\ell$ vectors of the fundamental quasiparticles. In the standard hierarchy, the number of fundamental quasiparticles grows with the level of the hierarchy, {\it i.e.\/} with the rank of the $K$-matrix, and the null vectors (if they exist) belong to the physical subspace. In contrast, in our approach, the physical subspace is spanned by the fundamental quasiparticle, whose $\ell$-vector is $\ell=(0,0,0,1)$. Thus,  although  in all the cases we have discussed  the $K$-matrices of the form of Eq.\ \ref{K} {\em have} null vectors, they  {\em do not} belong to the physical subspace. In other terms, in this construction the ``null-particles" are unphysical, and as a result these FQH states {\em are} $T$-stable. In particular, the observed state at $\nu=4/11$ is $T$-stable~[\onlinecite{T}].

Finally we note that this hierarchical construction, just as in any
hierarchy, at least at a formal level can be repeated an indefinite
number of times, leading to the construction of a (hierarchical) FQH
Jain state for any odd-denominator fraction. We will spare the reader from this discussion.

\section{Application to the observed states and the stability of the hierarchical Jain states}
\label{criterion}

In this section we show  that the states reported in Ref. [\onlinecite{pan03}], as well as some other states 
observed earlier on [\onlinecite{goldman90,du94}], can be reproduced  within this
framework. 

It is clear from the experiments that all  these states are not equally
stable, and that some criterion must be devised to classify these states. We will address this problem 
here at a very qualitative (and perhaps naive) level.

In the case of the primary Jain sequences the stability of the
states is determined by the size
of the excitation gap and, to a lesser extent, by the interaction coupling constants between 
quasi-particles. Here too a similar sort of stability criterion can be
constructed, by focusing mainly on the quasi-particle (and quasi-hole) energy gaps. 
We should note that an argument of this sort can at best  determine the relative stability 
for states sharing the same parent
Jain state. In particular, this sort of analysis should fail if a competing state of a different 
type were to be allowed for the same filling factor.
Another factor that must be taken into account is that  the states can be FQH states of either {\em
quasi-particles} or {\em quasi-holes}, which in general {\rm are not}
related simply by a particle-hole transformation as they have
different gaps  as well as interactions, and particle-hole symmetry holds only approximately. 

A related issue is that in
our construction of the hierarchical states there is more than one way to reproduce a given filling fraction. 
For instance the $\nu =4/11$ state can be
obtained as a $1/3$ descendant with $\nu_1=1/3$, or alternatively,
as a $2/5$ descendant with $\nu_1=-2/3$. Likewise, the $\nu=5/13$ state can
be constructed as  a
$1/3$ descendant with $\nu_1=2/3$, or as  a $2/5$ descendant with
$\nu_1=-1/3$.
In the case of $\nu=4/11$ the choice seems straightforward. It 
should be a  descendant of the more stable parent state ($1/3$) with the
smallest possible number of extra quasi-particles condensing into a
$\nu_1=1/3$ state. However, for the $5/13$ state it is not clear
whether the correct option is to choose the more stable parent state
($1/3$) with a larger number of condensing quasi-particles into a
$\nu_1=2/3$ state, or the less stable parent state ($2/5$) but with a
smaller condensate of quasi-holes into a $\nu_1=-1/3$. 
In the case of exact particle-hole symmetry these two constructions lead exactly 
to the same state even though they have different parent states. Notice that even 
though there are different ways to construct a given hierarchical state, the resulting 
state has exactly the same universal properties. 
Hence these are not distinct FQH states. 

In what follows, we will discuss two different criteria which are both intuitively reasonable and 
are natural within the framework of this mean field theory. A more
careful study of the energy gaps and of the interactions between the elementary excitations
(which is beyond the scope of this work)
is necessary to determine which of the following criteria is more suitable.

The first criterion is based on the following assumptions:
\begin{enumerate}
\item
The more stable parent states are the Jain states
with larger gaps. 
\item
The more stable hierarchical  states are those that
having the largest gaps, are descendants of the more stable Jain
states. 
\item
If there are two different ways of constructing a hierarchical
state, the
more stable one is the one that has the more stable parent state, and
that requires the lowest density of condensing quasi-particles. 
\item
Finally, states
with $\nu_1>0$ are more stable than those with $\nu_1<0$, since
quasi-holes are expected in general to have smaller excitations energies than quasi-particles. 
\end{enumerate}
It is apparent that this a rather qualitative criterion at best, and that in a number of cases it does not provide
for a unique answer. For instance, it is not obvious whether it is better to construct a state with quasi-particles 
(with $\nu_1>0$) of a more stable primary state, or with quasi-holes (with $\nu_1<0$) of a less stable primary 
state. Clearly it is not possible to determine a unique choice without some knowledge of the quasi-particle and
quasi-hole excitation energies (their gaps) and of their interactions [\onlinecite{order}]. 
Nevertheless it is still useful to explore what consequences follow from the naive application of this criterion.
 
According to this criterion we must first look at the the states which share the same parent primary Jain state:
\begin{itemize}
\item
{\it $1/3$ descendants}: they are obtained by setting  $p=1$ and 
$2n=2$, and are given by the sequences 
\begin{equation}
\frac{1}{\nu}=2+\frac{1}{1+ \nu_1}
\label{1-3}
\end{equation}
yielding states at filling factors $4/11$ (with  $\nu_1=1/3$), $5/13$
(with $\nu_1=2/3$) and $6/17$ (with $\nu_1=1/5$). 
Other descendants,
not seen yet in experiment, have 
filling factors $7/19$ ($\nu_1=2/5$), $10/27$ ($\nu_1=3/7$), $8/21$ ($\nu_1=3/5)$, and so on
\item
{\it $2/3$ descendants}: they are obtained by setting  $p=-2$ and 
$2n=2$, and are given by the sequences 
\begin{equation}
\frac{1}{\nu}=2+\frac{1}{-2+ \nu_1}
\label{2-3}
\end{equation}
According to our criterion, the most stable descendant states will be
$5/7$ (with $\nu_1=1/3$), $9/13$ (with $\nu_1=1/5$), $4/5$ (with
$\nu_1=2/3$), $8/11$ (with $\nu_1=2/5$), etc.
All these states fall outside of the range of filling fractions of the data of Ref. 
 [\onlinecite{pan03}], since they have $\nu >2/3$. However, a weak signature of a possible state at $5/7$ and
$4/5$ was observed earlier by Du {\it et al.\/} [\onlinecite{du94}] who saw a weak depression of the
longitudinal resistivity for these filling factors. Also, Goldman and Shayegan [\onlinecite{goldman90}] saw FQH states 
at $\nu=9/13$ (and at $7/11$).
\item
The  $2/7$  descendants  are obtained by setting  $p=-2$ and 
$2n=4$, and are given by the sequences 
\begin{equation}
\frac{1}{\nu}=4+\frac{1}{-2+ \nu_1}
\label{2-7}
\end{equation}
The states seen  in the experiment  have filling factors at  $4/13$ (with 
$\nu_1=2/3$), and $5/17$ (with $\nu_1=1/3$). 
\item 
The  $3/5$ descendants  are obtained by setting  $p=-3$ and 
$2n=2$, and are given by the sequences 
\begin{equation}
\frac{1}{\nu}=2+\frac{1}{-3+ \nu_1}
\label{3-5}
\end{equation}
So-far the  only observed  $3/5$ descendant has filling factor  $7/11$ (with 
$\nu_1=2/3$). 
\end{itemize}

A somewhat different classification of the observed states can be
obtained if we adopt  another stability criterion. In this second criterion we will assume instead that 
the more stable states are those generated
by the condensation of the smallest number of elementary excitations  on top of
the more stable Jain states, independently of whether they are
quasi-particles or quasi-holes. This is a very crude approximation, but
 consistent with our derivation of the quasi-particle condensate, since
except for the sign of the charge, it does not distinguish between quasi-particles and quasi-holes.

It is straightforward to show that given a parent state with filling
fraction $1/(2n+1/p)$, the number of elementary excitations on top of
it that condense into the state $\nu={\frac{p(2n_1p_1 +1)+p_1}{|\det K|
}}$ is given by 
$n_1^{qp}= 
{\frac {B}{2\pi}} {\frac{p_1}{|\det K| }}$ ( in units of $e=c=\hbar=1$). 
Therefore, if there are two
different ways of generating  a given filling
fraction, the more stable one will be the one that has the smallest
number of condensing excitations,  {\it i. e.\/}, the smallest $p_1$.
According to this criterion, for each Jain state, the more stable
descendants, and therefore the most likely to be observed,  will be
those with $\nu_1=\pm 1/(2n_1+1)$ ( {\it i. e.\/}, $ p_1=\pm
1$). We will see below that all the states generated following this
rule   have been
seen in reference [\onlinecite{pan03}].

Thus, following this second criterion we get instead the following classification:
\begin{itemize}
\item
{\it $1/3$ descendants}: 
In this case, the more stable descendant states will be 
 $4/11$ (with  $\nu_1=1/3$),  $6/17$
(with $\nu_1=1/5$), and  $4/13$
(with $\nu_1=-1/5$) . For $\nu_1=-1/3$ we obtain $\nu=2/7$ which is of
course more stable as a Jain (parent) state. 
The next states that can be obtained are $8/23$ ($\nu_1=1/7$), and
$6/19$ ($\nu_1=-1/7$), whose gaps are smaller than the ones observed by Pan {\it et al.\/} [\onlinecite{pan03}].
\item
{\it $2/3$ descendants}:
This yields the states $5/7$ (with $\nu_1=1/3$) , $7/11$ (with
$\nu_1=-1/3$), $9/13$ (with $\nu_1=1/5$), $11/17$ (with $\nu_1=-1/7$),
etc.
The state $9/13$ lies outside the range of the data of Ref. [\onlinecite{pan03}] (since $9/13>2/3$),
 although it was reported earlier in Ref. [\onlinecite{goldman90}]. However,
$11/17$ is within the range of the observed states, and in principle 
should be as stable as  $5/17$ or $6/17$, except for the fact that
it is the result of a quasi-hole condensate.
\item
{\it $2/5$ descendants}:
This yields the states $3/17$ (with $\nu_1=1/3$), and  $5/13$ (with
$\nu_1=-1/3$). Notice that $3/17$ being smaller than $ 2/7$ is out of
the range of the states observed in [\onlinecite{pan03}].
The next possible states  $5/27$ (with $\nu_1=1/5$), $9/23$ (with $\nu_1=-1/7$),
etc, have very small gaps.
The state $5/13$ can also be constructed as a $1/3$ descendant with
$\nu_1=2/3$, but this choice requires a larger number of condensing
excitations than the one obtained as a $2/5$ descendant.
\item
{\it $2/7$ descendants}:
  have filling factors  $5/17$ (with 
$\nu_1=1/3$),  $7/25$ (with $\nu_1=-1/3$), and states with even
smaller gaps.
\item
{\it $3/5$ descendants}:
  have filling factors  $8/13$ (with 
$\nu_1=1/3$),  $10/17$ (with $\nu_1=-1/3$), and states with even
smaller gaps.
As discussed before, the state $7/11$ can be constructed as a $3/5$
descendant with $\nu_1=2/3$, but this requires a larger number of
quasi-particles than the construction as a $2/3$ descendant.
\end{itemize}
\noindent

As mentioned before, besides the very general arguments invoked about
the stability of the  parent states, together with the size of the gap and the density of
condensing excitations, there are no other factors that one can take
into account at this level of approximation to decide on how to organize the states according to their degree of relative stability. 
A microscopic study of the interactions between
the quasi-particles and/or
quasi-holes involved is necessary to solve this question. Effective interactions among excitations of Jain states 
have been studied numerically recently  by Lee {\it et al.\/} [\onlinecite{jain02c}]. 
Goerbig {\it et al.\/} [\onlinecite{goerbig03}] have investigated the form of the quasi-particle interactions for 
Jain states with $\nu={\frac {1}{2n+1}}$ using the Hamiltonian approach for the fermionic
Chern-Simons theory of the FQHE of Murthy and Shankar [\onlinecite{shankar02}]. 

To conclude our discussion about the experimental results, we  turn our attention  to the minima in
$\rho_{xx}$, but not yet a plateau in $\rho_{xy}$, observed by Pan {\it et al.\/} [\onlinecite{pan03}] at the
even denominator fractions. We can construct these states as Jain descendants only if  $\nu_1=1/2$. 
In particular, we obtain   $3/8$ as a  $1/3$ descendant,  $3/10$  as a $2/5$ descendant, and  $5/8$
as a $3/5$ descendant. However, in this theory these
even-denominator states are {\em compressible} ~[\onlinecite{hlr}] fully polarized hierarchical Jain states since 
they are gapless. Alternatively, they have been  described as either  paired
states~[\onlinecite{read-moore,quinn03a,scajain}], or liquid crystal-like
states~[\onlinecite{jain-stripe,koulakov,chalker}].

\section{A partially spin polarized description of the observed states}
\label{partial}

Up to this point,  we have discussed how to  construct the FQHE states observed in Ref.
[\onlinecite{pan03}] as fully polarized  hierarchical descendants of the Jain series. Since this
 construction requires the condensation of the quasi-particles of the
 parent state, it is clear that these daughter states might become
 unstable under changes leading to a complete reorganization of the
ground state. Thus, if a FQH state of electrons were to become
available at the same filling  fraction, for instance a state which
involves the spin degrees of freedom, it will be necessary to
determine which state is chosen for a given physical system. Thus, the
application of pressure to the sample, or tilting of the magnetic
field may drive a transition to  a partially polarized, or
even an unpolarized state. 
Similar considerations apply to paired states.

In this section we will consider if partially polarized FQH states can
compete with the  hierarchical Jain states discussed in Section \ref{new}.
In  Ref. [\onlinecite{lopez01}] and Ref. [\onlinecite{lopez95}] we showed that
 the Chern-Simons field theory for two dimensional electron systems
 that have an extra degree of freedom, such as a layer or spin index,
 can describe FQHE states whose filling fractions are 
\begin{eqnarray}
\nu &=& {\frac{2n-(\frac 1 p_\downarrow +2n_\downarrow)-(\frac 1 p_\uparrow +2n_\uparrow)}{n^2- (\frac 1 p_\uparrow +2n_\uparrow)(\frac
1 p_\downarrow +2n_\downarrow)}}
\label{eq:ffspin}
\end{eqnarray}
Here, $2n_\sigma$ is the integer
 number of fluxes attached to the electrons with polarization 
$\sigma$ ($\sigma=\uparrow,\downarrow$), and $n$ is
 the integer number of fluxes attached to a given particle due to
 the presence of the particles with the opposite polarization. The
 integer $|p_\sigma|$ is the number of Landau levels filled by the
 particles with polarization $\sigma$. All these numbers can have either sign, since
  the flux attached, or the effective magnetic field seen at 
 mean field, can be either parallel or anti-parallel to the external magnetic field.

The total $z$-component magnetization per electron of the ground state
is
\begin{equation}
S_{total}^z = \frac {\nu_\uparrow -\nu_\downarrow}{2\nu}
\label{eq:pol}
\end{equation}

As discussed in Ref. [\onlinecite{lopez95}] and Ref. [\onlinecite{lopez01}], in the case of a spin-$1/2$
system, in order to preserve the $SU(2)$ spin rotation invariance of the Hamiltonian, even in the presence of a 
Zeeman term, we
must choose $2n_\sigma=n$. In other words, the flux attachment should be  done in such a way that
it does not distinguish between different spin  orientations. In
particular this guarantees that the statistics of all the
quasi-particles is the same, independently of their spin orientation. Notice that this is a {\em symmetry requirement} 
and it does not imply that the state is necessarily a spin singlet (although it is consistent with it).
It is simple to check that none of the states reported in Ref.
[\onlinecite{pan03}] can be obtained imposing the condition $2n_\sigma=n$,
$\sigma=\uparrow,\downarrow$.

On the other hand, if  we could assume that the physical system breaks the $SU(2)$ symmetry {\em explicitly}, 
we would be allowed to  relax this condition and only require that
the number of crossed flux ($n$) attached to the particles is even in order to maintain their  
fermionic statistics. In this case it is possible
to find realizations of the different observed states.
Taking for instance $n=2n_\uparrow$ and $2n_\downarrow-2n_\uparrow \equiv 2m$, we can
rewrite the total filling fraction in the following way
\begin{eqnarray}
\nu = \frac {1}{2n_\uparrow + {\frac {1}{p_\uparrow + \nu^*}}}
\label{eq:nupol}
\end{eqnarray}
where
\begin{equation}
\nu^* = \frac {1}{2m + {\frac {1}{p_\downarrow}}}
\label{eq:nupolstar}
\end{equation}
This expression looks identical to the one we found for the
hierarchical states (eq. (\ref{newnu2})) provided that 
$p_\uparrow \rightarrow p$, $2m
\rightarrow 2n_1$, $p_\downarrow \rightarrow p_1$, and 
$2n_\uparrow \rightarrow
2n$. 
However the interpretation of the integers appearing here is
different. In this case the flux attachment is performed on the
electrons, 
with
$2n_\uparrow=2n$ fluxes attached to electrons with spin $\uparrow$,
and $2n_\downarrow=2n +2n_1$ fluxes attached to electrons with spin
$\downarrow$. It is therefore obvious that the $SU(2)$ symmetry is
explicitly broken. Moreover,  all 
these states are  partially polarized with polarization 
$S_{total}^z = \frac {p_\uparrow (2m p_\downarrow +1) -
p_\downarrow}{2[p_\uparrow (2m p_\downarrow +1) +p_\downarrow]}$. On
the other hand, in the case of the hierarchical construction of
Section \ref{new},  $2n$ fluxes are attached to the original electrons
that condense into the parent state, and $2n_1$ fluxes to the
remaining excitations that condense into the daughter state.

It can be shown that the states with even denominator ($\nu=3/8,5/8$
and $3/10$) can not be
constructed as incompressible partially polarized ones if the condition $n=2n_\uparrow$ is imposed.
We could only obtain them if $\nu^*=1/2m$, but in this case,
they are compressible. Other possible mechanisms to generate even
denominator incompressible 
states involve pairing [\onlinecite{scajain,quinn03a}].

An alternative proposal 
has been put forth by Park and Jain [\onlinecite{parkjain}], who  discussed states whose
filling fraction is given by Eq. (\ref{eq:nupol}) for the particular
case of  $n=2,p_\uparrow=1$. The authors argue that these states are
mixed states of composite fermions of different flavors, carrying
different number of fluxes (vortices). In fact, the construction of Ref.  [\onlinecite{parkjain}] can be obtained from  
a scheme similar to the partially polarized states discussed in this section. Thus,
 we first attach $n=2n_ \uparrow$ fluxes to every
fermion, independently of the orientation of its spin. The remaining fluxes
$2n_\downarrow-2n_\uparrow=2m$ are attached only to the, let's say, spin $\downarrow$
electrons. 

However, as discussed above, a construction of this type, in which different spin 
orientations get attached different number of flux quanta, breaks explicitly the 
$SU(2)$ spin symmetry of the system, and  it is not an equivalent representation 
of the system. Naturally, it is always possible to carry out the formal process of 
flux attachment in an asymmetric fashion, {\it i. e.\/} by ignoring the fact that, except for the presence of the Zeeman term 
(which commutes with the Hamiltonian), the physical system is $SU(2)$ 
invariant. In such an approach, spin $SU(2)$ becomes a dynamical symmetry. However, as soon as approximations are 
made, the $SU(2)$ symmetry is explicitly broken. Of course, since the original problem has an $SU(2)$-symmetric Hamiltonian, 
the effects of the symmetry should eventually be recovered. However there is no guarantee that the symmetry can be recovered 
in perturbation theory, as it may well be a non-perturbative effect. Thus,  the resulting 
mean field theory of composite fermions is constructed with states
which are essentially orthogonal to the physical states, and the fully
projected states are very far from the intuitive and simple wave
functions of (composite) fermions filling up these effective
asymmetric Landau levels. Thus, although Park and Jain succeeded in constructing states of this type for most of the new fractions, 
the states they find are not fully polarized. Moreover, it is unclear how this scheme can be made compatible with the spin symmetry 
of the underlying Hamiltonian. We should emphasize that, to have a ground state which is either fully or partially polarized, it is not necessary to break the $SU(2)$ symmetry of spin at the level of the Hamiltonian (beyond the effects of a possible Zeeman term).  

\section{Edge states and electron tunneling}
\label{edges}

In this Section we give a brief derivation of the theory of the edge states for the hierarchical 
Jain states discussed in 
Section \ref{new}. Since we will follow closely the approach we used in Ref. [\onlinecite{lopez99}] and 
Ref. [\onlinecite{lopez01}] 
we will omit many details and refer the reader to these references. The main goal of this section is to determine the exponent of the differential tunneling conductance for electrons.

The Lagrangian of Eq.(\ref{eq:Ljain}) has the standard form discussed by Wen and Zee
[\onlinecite{wz}]. Therefore, following the general arguments of 
Ref. [\onlinecite{wen}], it is straightforward to extract a theory for the edge
states which reflects the structure of the bulk.
In what follows we assume that there is a sharp  potential
that confines the electrons to a disk in such a way that there is no
edge reconstruction. Furthermore we will also assume~[\onlinecite{lopez99,lopez01}] that the possible multiple edge charge 
modes are unresolved, {\it i. e.\/} that they are confined within a magnetic length. Of course the validity of this assumption 
depends
on microscopic details such as the specific form of the confining
potential.

The effective theory takes its simplest form when written in terms of the chiral
boson $\phi_C$, the charge mode, which  is the only mode that couples to the
electromagnetic field, and non-propagating topological
modes which play a role  similar to that of Klein factors.
In terms of these  fields, the
Lagrangian for the edge theory is
\begin{eqnarray}
{\cal L} &=&  {\frac {1}{4\pi \nu}} (\partial_1
\phi_C \partial_0 \phi_C - v  \partial_1
\phi_C \partial_1 \phi_C) \nonumber \\
&-& {\frac {1}{4\pi}} (p+\nu_1) \; \partial_1
\phi_{T} \partial_0 \phi_{T}\nonumber \\
&-& {\frac {1}{4\pi}} \kappa_{ij} \partial_1
\phi_{T_i} \partial_0 \phi_{T_j}
\label{eq:bdrydiag}
\end{eqnarray}
where $v$ the velocity of the charged  edge
 mode. The $2 \times 2$ matrix $\kappa_{ij}$ is given by
\begin{eqnarray}
\kappa =  
\begin{pmatrix}
 p  & 1    \\
 1 &  -2 n_1  
\end{pmatrix}
\label{Kappa}
\end{eqnarray}
A general edge operator can be written as
\begin{equation}
\Psi(x) = e^{i ( \alpha_C \phi_C + \alpha_T \phi_T + \sum_{i=1}^2 \alpha_{T_i}
\phi_{T_i})}
\label{eq:edgegral}
\end{equation}
The charge $Q$ and statistics $\theta$ of these excitations are 
\begin{eqnarray}  
{\frac Q e} &=&  - \nu \alpha_C \nonumber \\
{\frac {\theta}{\pi}} &=& - \nu \alpha_C^2 +  {\frac {1}{p+\nu_1}}
\alpha_T^2 +\alpha_{T_i}
\kappa^{-1}_{ij} \alpha_{T_j}
\label{eq:qnCT}
\end{eqnarray}
It is apparent  that the topological modes only
contribute to determine the statistics of the operators. Their role is to provide for  a 
set of effective Klein factors which give the physical excitations their correct statistics.

The  operator that creates a quasi-particle or quasi-hole at the
boundary can be found by requiring that
its charge and statistics are given by Eqs. (\ref{eq:newQ}) and
(\ref{eq:newtheta}) respectively. It is immediate to see that the
 quasi-hole operator can be written as 
\begin{eqnarray}
\Psi_{\rm qh}(x)=e^{i ({\frac{1}{p(2n_1p_1+1)+p_1}}\,  \phi_C +\phi_{T})}
\label{eq:qptcle}
\end{eqnarray}
Analogously, the electron operator at the boundary is
\begin{eqnarray}
\Psi_{\rm e}(x)=e^{i ({\frac{1}{\nu}} \phi_C +\ |\det K| \phi_{T})}
\label{eq:el}
\end{eqnarray}
This operator has the correct  charge   $Q= -e$,  and statistics  
$\theta /\pi = |\det K|(2n_2 (2np+1) + 2n)$, measured
with respect to fermions. Notice that 
all the physical operators can be represented only in terms of the
charge mode $\phi_C$ and one of the topological modes $\phi_T$.

There is much  interest in  studying different tunneling processes into the
edges of the FQHE hierarchical states. In  order to do so,   we need
to compute the propagator for an excitation created by an
operator of the form of Eq. (\ref{eq:edgegral}). Since the effective
action for the edge modes is quadratic in the fields, the calculation
of the  propagators
of the chiral bosons is
straightforward giving [\onlinecite{lopez99,lopez01}]
\begin{eqnarray}
<\phi_C(x,t) \phi_C (0,0)> &=& - \nu  \ln z \nonumber \\
<\phi_T (x,t) \phi_T (0,0)> &=&  i {\frac {\pi}{2(p+\nu_1)}} 
{\rm sgn} (xt)
\label{eq:propfields}
\end{eqnarray}
where $z=x +i v t$.

Using the above results, the propagator for the field $\Psi(x)$, of the form given in Eq. (\ref{eq:edgegral}), 
in the limit $x \rightarrow 0^+$ becomes
\begin{eqnarray}
<\Psi^\dagger (0^+,t) \Psi(0,0)> \propto {\frac {1}{|t|^{g_t}}} e^{i
{\frac {\theta}{2}} {\rm sgn}(t)}
\label{eq:propop}
\end{eqnarray}
where  $g_t = \nu \alpha_C^2$ and $\theta$ is given by
Eq. (\ref{eq:qnCT}). In particular, we find that the exponent $g_t$ for electrons is
$g_e=1/ \nu$, whereas for quasi-particles (and quasi-holes) $g_{\rm qp}= \nu / (p(2n_1p_1+1)+p_1)^2$.

The tunneling current $I$ at  bias voltage $V$ has the
scaling form [\onlinecite{weniv,kaneiv}] $I(V) \propto V^\alpha$, where the
exponent $\alpha$ is determined by the scaling dimension of the
tunneling operator. There are  three 
cases of physical interest: (a) internal tunneling of quasi-particles, for which
$\alpha_{qp}= 2 g_{qp} -1$ (here scaling holds at large bias $V$), (b)
tunneling of electrons between identical fluids, for which $\alpha_e= 2 g_e -1$ (at low bias $V$), and (c) electron
tunneling between distinct fluids, for which $\alpha_t = g_e$ (again at low bias $V$).
In particular, in the case of tunneling of electrons from a Fermi
liquid into a hierarchical  FQH state, we find that the tunneling exponent is $\alpha= g_e= 1/\nu$. 

Thus, in contrast with what is obtained for the Haldane-Halperin
hierarchy~[\onlinecite{wen}], all the  states described by our approach 
have a universal value of their electron tunneling exponent, which is
always equal to $1/\nu$ in the case of an unreconstructed unresolved edge.
Although this interesting result agrees with the experimental observations of references
[\onlinecite{chang96,grayson98,hilke01,hilke03}], we should note that it does
{\em not} predict that  the exponent is $1/\nu$ for {\em all} values
of the magnetic field, as the experiments seem to suggest. Instead
what this theory predicts is that for each FQH state at filling factor
$\nu$, the exponent will {\em lock} at the value $1/\nu$ so far as the
bulk state has not changed. 
Thus, there should still be a plateau, no matter how small, centered
about the value $1/\nu$.  Although this reproduces the trend seen in
the experiment, it still requires that the functional dependence of
the exponent on the magnetic field has the same structure as the Hall
conductance. 
The fact that, with the possible exception of a very small plateau at
$1/3$ with a small but observable discrepancy in the value of the
exponent~[\onlinecite{grayson98}], no plateau has ever been seen in edge
tunneling exponent  remains an unsolved and intriguing puzzle which 
presumably will not be resolved until a true point-contact geometry 
becomes available for experiments.

\section{Conclusions}

In this paper, we presented an effective theory to describe
hierarchical states constructed as daughter states  of
the principal fractional quantum Hall Jain
series at $\nu= {\frac {p}{2n p+ 1}}$. These daughter states result
from the condensation of the elementary excitations (either quasi-particles or
quasi-holes) of the parent (Jain) state, into new FQH Jain-like states. This
effective theory is a generalization of  the fermionic Chern-Simons theory of the FQHE for
fully polarized systems~[\onlinecite{lopez91,lopez99,lopez01}]. It reproduces all
the odd denominator filling fractions reported in Ref. [\onlinecite{pan03}] as
well as some states reported earlier in Refs. [\onlinecite{goldman90,du94}]. 
We found that these states are both locally stable and $T$-stable.

Our results do not preclude the existence of other possible competing states, such as paired states, 
inhomogeneous states (stripes, bubbles, etc.) or anisotropic nematic states. But they do confirm that at least 
locally (in energy space) the hierarchical states are stable states of the 2DEG.

We  also used a similar approach to discuss the possibility of explaining some of these states
as partially polarized FQH states. We  found that it is not possible to
construct the  new states reported in Ref. [\onlinecite{pan03}] using a scheme which 
respects the $SU(2)$ invariance of the underlying Hamiltonian.

 Finally, we 
derived  a theory of the edge states for the new hierarchical
states, using the Chern-Simons bulk effective theory presented
here. In particular we discussed the predictions of this theory for
the edge tunneling exponent of electrons from a Fermi liquid, and find
that for an unreconstructed sharp unresolved edge, the tunneling
exponent is always equal to $1/\nu$ for all the fully polarized Jain states (hierarchical or not). 

\begin{acknowledgments}
We thank W. Pan for useful discussions. This work was supported in
part by the NSF grant DMR 01-32990 (EF), and by CONICET, Argentina (AL).
\end{acknowledgments}


\begin{thebibliography}{50}
\bibitem[{not()}]{note}
\bibinfo{note}{{Current state-of-the-art finite size numerical diagonalizations
  are done on clusters with up to 24 electrons in the lowest Landau level. For
  such large clusters, finite size effects should be small for filling factors
  in the primary Laughlin and Jain states. However, for hierarchical states a
  large fraction of the electrons are effectively ``used up" in making up the
  underlying primary FQH state, and relatively few electrons (about seven) are
  left to make quasiparticles which should eventually condense. Thus,
  hierarchical states are expected to have significantly larger finite size
  effects compared to their parent primary states.}}
\bibitem[{foo({\natexlab{b}})}]{foot-edge}
\bibinfo{note}{{Naturally, this result means that the exponent has a plateau in
  the value of the exponent, no matter how small, whenever the bulk is in a FQH
  state with a given Hall conductivity.}}
\bibitem[{gap()}]{gap-estimate}
\bibinfo{note}{{Estimates of these interactions for just two quasi-particles
  above the $\nu=1/3$ Laughlin state have been calculated for small clusters of
  electrons, and thus with presumably significant finite-size effects, in Refs.
  [\onlinecite{quinn03a,quinn03c,quinn03b}]. These results suggest that the gap
  is significantly smaller than the mean field estimate of Eq. (\ref{gap}), and
  that in certain models it may even vanish. In fact, it has been suggested
  that the new FQH state at $\nu=4/11$ is either partially polarized
  [\onlinecite{jain02a,jain03b}] or paired [\onlinecite{quinn03a,quinn03b}].}}
\bibitem[{T()}]{T}
\bibinfo{note}{{The state at $\nu=4/11$ appears already at the first level of
  the Halperin-Haldane hierarchy where it results from the condensation of
  multiple vortices (or quasiholes) of the parent $1/3$ state. Using the
  notation of Refs. [\onlinecite{haldane,halperin,wen}], it is constructed as
  the state with $p_1=3$ and $p_2=4$. In this sense it is less stable than the
  $2/5$ state which involves the condensation of quasiholes of the parent $1/3$
  state, and has $p_1=3$ but $p_2=2$. Notice that in both cases the $K$-matrix
  is $2 \times 2$ and does not have null vectors. Thus, contrary from what a
  cursory reading of Ref. [\onlinecite{haldane-t}] may suggest, the state at
  $4/11$ is $T$-stable.}}
\bibitem[{ord()}]{order}
\bibinfo{note}{{Although in most situations the quasi-hole gap is expected (for
  the same state) to have a smaller gap than a quasiparticle, there are model
  interactions for which the order may be reversed. However this does not
  appear to be the case in any of the published studies. See, for instance,
  Refs. [\onlinecite{jain02c,goerbig03}].}}
\end{thebibliography}


\end{document}